%% file: paper.tex
\documentclass{sig-alternate-05-2015}

\usepackage{cite}

\usepackage{algpseudocode}
\usepackage{algorithm}
\usepackage{times}
\usepackage{amssymb}
\usepackage{amsfonts}
\usepackage{balance}

\usepackage{graphicx}
\usepackage{caption}
\usepackage{subcaption}

\usepackage{pifont}

\newcommand{\xmark}{\ding{55}}

\algnewcommand\And{\textbf{ and }}
\algnewcommand\Or{\textbf{ or }}

\newdef{definition}{Definition}
\newtheorem{theorem}{Theorem}
\newdef{example}{Example}

\begin{document}
%

\title{Pushing the Limits of Encrypted Databases with Secure Hardware}

%
%
%
%
%

%

\author{ 
\normalsize
\begin{tabular}{c}
\begin{tabular}{ccccc}
Panagiotis Antonopoulos &  Arvind Arasu &  Ken Eguro &  Joachim Hammer    \\
Raghav Kaushik & Donald Kossmann & Ravi Ramamurthy & Jakub Szymaszek 
\end{tabular} \\
\rule{0ex}{3ex} Microsoft Corporation \\
{\small \texttt{ \{panant,arvinda,eguro,johammer,skaushi,donaldk,ravirama,jaszymas\}@microsoft.com }}
\end{tabular}
}

\maketitle

\input{abstract}
\input{intro}
\input{typesystem}

\input{attacks}
\input{reasoning}
\input{query}

\input{updates}

\input{related}
\input{concl}

\bibliographystyle{abbrv}
\bibliography{security}  

\end{document}

%% file: abstract.tex
\begin{abstract}
Encrypted databases have been studied for more than 10 years and are
quickly emerging as a critical technology for the cloud. The current
state of the art is to use property-preserving encrypting techniques
(e.g., deterministic encryption) to protect the confidentiality of the
data and support query processing at the same time.  Unfortunately,
these techniques have many limitations. Recently, trusted computing
platforms (e.g., Intel SGX) have emerged as an alternative to
implement encrypted databases. This paper demonstrates some
vulnerabilities and the limitations of this technology, but it also
shows how to make best use of it in order to improve on
confidentiality, functionality, and performance.  
\end{abstract}

%% file: intro.tex
\section{Introduction}

After more than 10 years of research \cite{Mehrotra:SIGMOD2002},
encrypted databases are becoming a reality. There are a number of
promising start-ups in this space (e.g., Privic, Cryptonor, and
ZeroDB) and, among established vendors, Microsoft recently shipped the
Always Encrypted feature for both the on-premise and cloud versions of
SQL Server \cite{sql-AE}. CryptDB \cite{CryptDB} is a famous research
system that has become iconic for the current popularity of encrypted
database systems.

The goal of an encrypted database is to maintain {\em confidentiality}
of the data. That is, the data should only be readable by the owner of
the data and other entities who have been authorized by the owner of
the data.  The data should not be accessable to hackers who have gained
control of the machine that hosts the database.  In a cloud scenario,
the data should not be accessable to the cloud provider, its
administrators or to co-tenants who run other applications on the same
machines.  All these guarantees should be maintained under the
strongest possible assumptions, including security vulnerabilities of
the operating system and other libraries on the machine. At the same
time, the goal is to run as much database {\em functionality} as possible
(ideally, the full SQL standard and other modern features) and at the
same cost and performance as a regular, non-encrypted database system.

\begin{figure}
\centering{
\includegraphics[width=0.33 \textwidth]{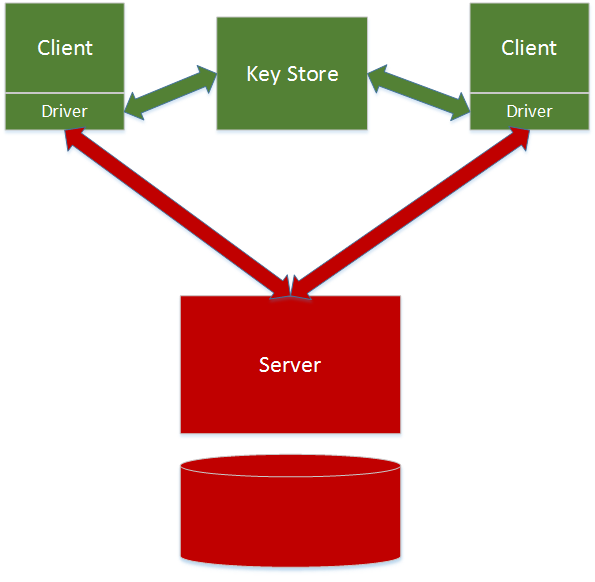}
}
  \caption{Client-side Encryption}
  \label{fig:cse}
\end{figure}

To achieve these goals, all encrypted database systems have adopted an
architecture that is called {\em client-side encryption}, depicted in
Figure \ref{fig:cse} \cite{Mehrotra:SIGMOD2002, Agrawal:SIGMOD2004,
  CryptDB, Sanamrad:DBSec2014}.  Data are encrypted on client machines,
which are assumed to be secure and have access to all encryption keys.
The encrypted data are sent from clients to the database server that
stores the data persistently.  
Queries are issued by applications at the client and are
rewritten by a DB driver to encrypt query constants. The rewritten,
encrypted queries are then sent to the server and executed. The server
returns encrypted query results, which are in turn decrypted by the DB
driver at the client. 

To effect queries on encrypted data on the (untrusted) database
server, the current generation of encrypted database systems leverage
{\em property-preserving encryption (PPE)}; this approach is often
also referred to as partially homomorphic encryption.  The key idea
of PPE is that, depending on the encryption algorithm used, the
database system can carry out operations upon the
encrypted data without decrypting it. A prominent example is {\em
  deterministic encryption} (e.g., AES in ECB mode), which produces
the same ciphertext when presented with the same plaintext; i.e., $x =
y \Leftrightarrow \mbox{\em enc}(x) = \mbox{\em enc}(y)$ for two
plaintext values, $x$ and $y$, and a deterministic encryption
function, {\em enc()}.  As a result, the database system can evaluate
equality predicates (point queries), equi-joins (e.g., foreign-key /
primary key joins), and group-by operations. Another example is {\em
  order-preserving encryption} (OPE); e.g.,
\cite{Agrawal:SIGMOD2004, Boldyreva:OPE2012, Sanamrad:DBSec2014,
  Popa:OPE2013}. OPE has the following property: $x \leq y
\Leftrightarrow \mbox{\em enc}(x) \leq \mbox{\em enc}(y)$. With OPE,
the database system can execute range predicates, Top N, and order-by
operations directly on encrypted data.  Finally, {\em Paillier
  encryption} \cite{DBLP:reference/crypt/Paillier11} is a way to
support arithmetics on encrypted data.  Microsoft Always Encrypted is
based on deterministic encryption \cite{sql-AE} and CryptDB supports a
variety of different PPE techniques \cite{CryptDB}.  

Unfortunately, the state-of-the-art in property-preserving and
homomorphic encryption has many limitations. PPE such as deterministic
encryption, OPE, or Paillier is good to support a specific
functionality; however, it does not compose. That is, there is no PPE
scheme that supports both arithmetics (as with Paillier) and
comparisons (as with deterministic): It is either one or the other so
that it is not possible to process predicates of the form $A + B = C$,
if $A$, $B$, and $C$ are encrypted.  Furthermore, there is a great
deal of SQL functionality (e.g., LIKE predicates with pattern
matching) for which no good PPE technique is known.  Finally, PPE
techniques often do not preserve SQL semantics for errors; for
instance, Paillier does not handle overflow errors correctly.
Another problem of many PPE techniques is that they leak information,
thereby compromising confidentiality for functionality
\cite{Seny:PPE2015}. Full homomorphic encryption \cite{Gentry:FHE2009}
overcomes these problems: It is general, can implement any circuit,
and it is semantically secure.  However, to date, full homomorphic
encryption is impractical from a performance perspective.    

\begin{figure}
 \centering
\includegraphics[width=0.33 \textwidth]{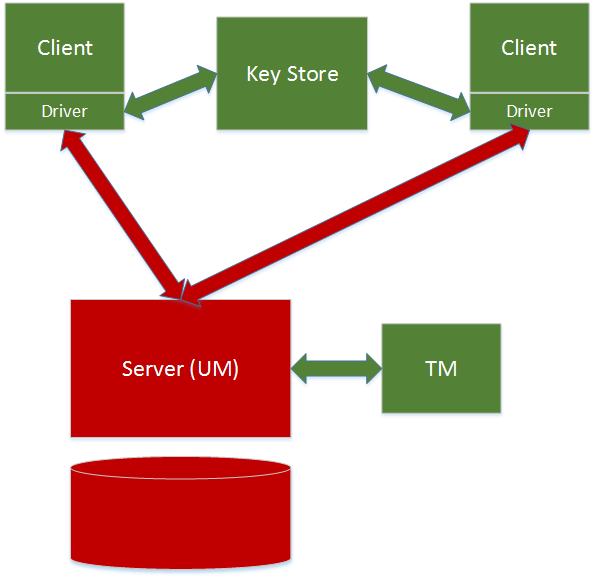}
  \caption{Client-side Encryption with Server-Side Trusted Machine}
  \label{fig:umtm}
\end{figure}

Recent advances in trusted computing platforms have given rise to an
alternative way to implement encrypted databases. These trusted
computing platforms promise to provide strong security features and
support arbitrary computation.  IBM's secure co-processor
\cite{ibm-coprocessor}, Intel SGX \cite{SGX:Explained}, and FPGAs
\cite{Eguro:FPL2012} are all based upon secure hardware which provides
secure storage and compute capabilities.  Similarly, special
hypervisor extensions such as VSM \cite{vsm-security} are designed to
provide stronger isolation and code integrity guarantees.  Figure
\ref{fig:umtm} shows the extended encrypted database architecture that
makes use of this technology.  Here, the server is enhanced with a
trusted computing platform that acts as a secure co-processor and runs
a special process that we call the {\em trusted machine (TM)}.
Clients grant the TMs access to their encryption keys (through secure
protocols), and the TMs can then compute any operation on the
encrypted data by decrypting the data, carrying out the operation, and
then encrypting the result. Projects that have explored this
architecture are TrustedDB \cite{TrustedDB:SIGMOD2011}, Cipherbase
\cite{Cipherbase:CIDR2013,   Cipherbase:ICDE2015}, Haven
\cite{Haven:OSDI2014}, and VC3 \cite{VC3:IEEE2015}.

Such trusted computing platforms do not have the limitations of PPE or
full homomorphic encryption:  They are Turing-complete and have acceptable
performance.  For instance, such TMs can
compose functionality to support comparisons and arithmetics on the
same encrypted data.  Furthermore, they are reasonably secure:  There
are known side-channel attacks for Intel SGX \cite{SGX:Explained},
but the technology is only going to get better.  Despite all these
features, it is nevertheless challenging to build an encrypted
database system using trusting computing platforms.  As we will show
in this paper, if not done correctly, all the security advantages of
trusted computing platforms are gone and attackers can easily get
access to confidential and encrypted data.  Furthermore, there are
still some basic database features such as integrity constraints and
certain kinds of updates that cannot be implemented without leaking
information even when using such trusted computing platforms and
assuming that they are secure.

This paper makes three contributions.  First, it generalizes and
formalizes the definition of an encrypted database system whose
implementation is based on a trusted computing platform (Section
\ref{sec:typesystem}).  Based on this formalization,
our second contribution is to demonstrate novel attacks on encrypted
database systems with trusted computing platforms that have not been
studied before (Section \ref{sec:attacks}).  Given these attacks, it
becomes clear that it is not possible to build a {\em perfect}
encrypted database system that leaks no information on confidential
data, implements the full SQL standard, and has good performance for
any SQL query, even assuming that the trusted computing platform is
fully secure and not subject to side-channel attacks, bugs, or other
vulnerabilities. Based on this observation, our third and arguably
most important contribution is a clean characterization of different
levels of confidentiality.  Specifically, we study a scheme to
implement an encrypted database that leaks no information, is able
to process all read-only queries, but has both functional and
performance limitations. Furthermore, we study relaxed schemes that
have good performance and implement (almost) the full SQL standard, but are  
subject to potential inference attacks.  While non of these schemes
are perfect, they dominate state-of-the-art encrypted systems
that are based on PPE (e.g., CryptDB and Always Encrypted) in all
regards: functionality, performance, and confidentiality.  

%% file: typesystem.tex
\section{Encrypted Databases}
\label{sec:typesystem}

This section gives a formal definition of an {\em encrypted
  database}.  This definition generalizes the definitions used in
related projects such as CryptDB \cite{CryptDB} and Microsoft Always
Encrypted \cite{sql-AE}.  This definition helps us to reason about
attacks on encrypted databases and different levels of
confidentiality (Sections \ref{sec:attacks} to \ref{sec:updates}).

Encrypted databases allow users to specify
which data is confidential and needs to be encrypted and which kinds
of operations are supported on which data.  For instance, {\em credit card
numbers} are typically declared to be confidential; equality is often
needed to validate a credit card number, but range predicates and
arithmetics are rarely needed on credit card numbers.  As another
example, {\em salaries} are also confidential in many applications,
but often require a broader set of operations, including arithmetics;
e.g., raise a salary by 10 percent.

In systems that are based on PPE (like CryptDB and Always Encrypted),
the encryption method implicitly determines the set of operations.  In
systems that make use of trusted computing platforms, the set of
operations on an encrypted column needs to be specified explicitly.
As will become clear in Sections \ref{sec:attacks} and later, it is
exactly this specification of the functionality that determines the
confidentiality, performance, and functionality of an encrypted
database.  It will also become clear that there are many ways to
effect different functionality and that unfortunately, it is easy
to get wrong.

\subsection{Confidential Data Types}
\label{subsec:type_types}

Like any other relational database, an encrypted database
is a collection of tables.  Each table has a schema that specifies the
name of the table and the types of all the columns of the table.  What
makes an encrypted database special is that the type of a column can
be a {\em confidential type} (or {\em encrypted type}), indicating
that the values of this column need to be encrypted.


In general, a type of an encrypted SQL database is defined by its {\em
  domain}, its {\em encryption method}, and a {\em key} to encrypt
values of that type. More formally, a type is a triple, $\{{\cal D},
{\cal E}, {\cal K}\}$.   

\paragraph*{Domain} Just as in any other programming environment, the
domain of a type defines all the values that an instance of that type
may have.  In SQL, the domain of a type may include {\em null}
values.  Furthermore, we encode errors as special values of the domain
to better compose functions in the event of errors (Section
\ref{sec:query}).  For instance, the {\em Integer} domain includes
error values that represent the result of ``1 / 0'' (division-by-zero
error) and overflows.  

A domain is \emph{length indistinguishable} if any two values in the
domain have the same representational length. A domain is \emph{length
  bounded} if there exists an upper bound $M$ for the representational
length of any value in the domain. For example, \texttt{char(n)} is
length indistinguishable, while \texttt{varchar(n)} is length
bounded. A length bounded domain can be made length indistinguishable
by padding. Length indistinguishability is important for formal
confidentiality statements because standard encryption schemes such as
AES on strings reveal the length and unintended information might be
inferred from this length in many applications.  

\paragraph*{Encryption Technique} In theory and for all of the
discussions in this paper, two encryption algorithms are sufficient:
(a) ``probabilistic'' (e.g., AES in CBC mode) encryption that is
\emph{IND-CPA secure (indistinguishability against chosen plaintext)}
and (b) ``plaintext'' (i.e., not encrypted at all for non-confidential
data).  Any other encryption algorithm can be simulated in our type
system with the help of rules (Section \ref{subsec:type_rules}).  In
practice and for performance reasons, however, we will also consider
PPE algorithms such as deterministic encryption (Section
\ref{subsec:query_deterministic}). 

\paragraph*{Encryption Key} The third component that defines a type is
the encryption key. In practice, 256-bit AES keys are common. For
plaintext, no key is needed; i.e., $T.{\cal K} = \emptyset$ if
$T.{\cal E} =$ {\em plaintext}. 

\subsubsection{Naming Conventions}

In the remainder of this paper, we use the following naming
conventions for types. First, we use {\em Table.column} or just {\em
  column} (if the table is clear) to denote the type of a specific
column. For instance, {\em Employee.name} denotes the type of the {\em
  name} column of the {\em Employee} table.  

Second, we often use ``one-time-keys'' to encrypt query constants or
intermediate query results. One-time-keys are encryption keys that can
be used only for a limited duration of time (e.g., in the context of
processing a single query) and may not be used to encrypt any data
that is persistently stored in the database. We denote types that are
based on one-time-keys using the {\em Temp} keyword. For instance,
{\em Temp(Employee.name)} denotes a type that has the same domain as
{\em Employee.name}, uses a one-time-key which is different from the key
of {\em Employee.name}, and may also differ in terms of the encryption
technique from {\em Employee.name}. {\em Temp(Integer)} denotes a type
that has the domain Integer, uses a one-time-key which is different
from any other key used in the system, and has a custom-defined
encryption technique. If the original type is unambigious, we just use
{\em Temp} to denote a derived type with a one-time key.

Finally, we use the traditional names to denote plaintext types. For
instance, {\em Integer} denotes the traditional, unencrypted Integer
type, including the {\em not-a-number (NAN)} and {\em null} values for
integers. Likewise, {\em Boolean} denotes the traditional, unencrypted
Boolean type of SQL with its three-value logic.

\subsection{Rules}
\label{subsec:type_rules}

We uses rules over types to control which operations are
permitted over which types in an encrypted database. A rule is of the form:  

\[ f (T_{i1}, \ldots, T_{in}) \rightarrow \langle T_{o1}, \ldots, T_{om} \rangle\]

Here, $f$ is an $n$-ary function with domain $T_{i1}.{\cal D} \times
\cdots \times T_{in}.{\cal D}$ and range $T_{o1}.{\cal D} \times \cdots
\times T_{om}.{\cal D}$. Such a rule specifies that the TM can execute
$f$ over the inputs encrypted as specified by the input types and
produces its outputs encrypted as specified by the output types. For
example, we could provide rules for basic functions such as ``add'',
``divide'', or ``equal''. A rule, however, can also specify more
complex functions, predicates, and even user-defined functions.  For instance, to evaluate
predicates of the form $A + B = C$ on three encrypted integers, we
could define the following rule: {\it addEqual(EncryptedInteger,
  EncryptedInteger, EncryptedInteger) $\to$ EncryptedBoolean}.  

For encrypted database that are based on PPE, the rules are defined
implicitly; i.e., {\em equal(T, T) $\to$ Boolean}, if $T$ is
deterministically encrypted. In our more general setting that makes
use of TMs, it is important to define and reason about all rules
explicitly. 

\subsection{Schemas with Type Metadata}
\label{subsec:schema}

To integrate our type system into a relational database system, we
propose to extend the SQL DDL in the following ways: 
\begin{itemize}
\item Declare and specify all the types used in the database.
\item Specify all the rules. Rules may only refer to declared types or
  {\em temp} types with one-time-keys derived from declared types. 
\item Associate each column of a table with a declared type.
\end{itemize} 

We refer to the types declared in such a schema as {\em schema types}.
In contrast to ordinary types which are defined in Section
\ref{subsec:type_types}, schema types are associated to a rule set.
In the context of a database and a rule set, it is possible to reason
about the information leakage and strength of a type.  For instance,
we can have two types: one for credit card numbers and one for
salaries. Both types could use AES encryption.  However, we likely
want to have stronger security guarantees for credit card numbers and
we accomplish this by providing a richer set of rules 
that apply to salaries (e.g., a ``percent-raise'' function) than for
credit card numbers.  Thus, the context (rule set and database)
determines the strength of a type and Section \ref{sec:reasoning}
formalizes this property.   Since the focus of this paper is on
encrypted database systems, we will mostly use the term {\em type} in
the remainder of this paper, even though strictly speaking we should
be using the term {\em schema type} to refer to a type in the context
of a database with a rule set.   

Conceptually, the owner of a database can freely specify a rule set to
their liking. We propose, however, that a database system
offers predefined rule sets for different levels of confidentiality
because it is easy to get wrong (and thus, insecure). Section
\ref{sec:query} discusses examples of such rule sets that correspond
to levels of confidentiality that are used in practice today.  

Several columns of the same or different tables can be associated to
the same type. That is, there is an N:1 relationship between columns
and types.  For instance, often the primary key column and all foreign
key columns that refer to that primary key have the same type. For
brevity and without loss of generality, this paper only discusses
scenarios in which all values of a column are encrypted using the same
key; i.e., a column is associated to only one and not multiple types.
Although support for multiple encryption keys and types within a
column is useful for multi-tenant database systems
\cite{Jacobs:SIGMOD2008}, throughout this work we assume that such
multi-tenant databases are horizontally partitioned by tenant.

\subsection{Implementation}
\label{subsec:implementation}

There have been several projects that have implemented an encrypted
database system using a trusted computing platform.  For this work, we
adopted the approaches taken in the Cipherbase
\cite{Cipherbase:ICDE2015} and VC3 projects \cite{VC3:IEEE2015}.  In these
approaches, the computation is split between a trusted machine (TM)
and an untrusted machine (UM) as shown in Figure \ref{fig:umtm}.  The
UM carries out all I/O and communication with clients as no
state-of-the-art trusted computing technology is able to do 
I/O\@. Furthermore, the UM carries out all computations on unencrypted
(public) data or computations on data encrypted using PPE, if the
operation matches the PPE scheme (e.g., comparison and deterministic
encryption).  The TM is only needed to carry out operations on
encrypted data and if that operation requires decrypting the data. 

Query processing proceeds as follows:  The client rewrites a query and
encrypts all constants, as described in the introduction. The UM at
the server compiles and optimizes the rewritten query.  During
compilation, the server has access to all the meta-data (types and
rules), detects all operations that must be carried out by the TM, and
generates a corresponding query plan.  The UM, in 
turn, executes the query by interpreting the query plan and calling
the TM as needed and returns the (encrypted) query results to the
client.

Most of this query processing process is straight-forward and the same
as in a traditional (non-encrypted) database system
\cite{Cipherbase:ICDE2015}.  There are two critical steps in this
process, however:  First, before the TM can execute any operations on
encrypted data, it must receive the keys from the client.  We call
this process {\em program registration} and describe it below.
Second, query optimization heavily depends on the rules specified by
the developer or security officer.  For
instance, hash joins can only be used if there is a rule that allows
the TM to compute a {\em hash function} on the encrypted data.
Section \ref{sec:query} discusses how different rule designs impact
query optimization.

To give an example, consider the following query:
\begin{quote}
  SELECT * FROM Employee \\ WHERE department = ``IT'' and salary > 500;
\end{quote}
Assume that {\tt department} is not confidential (plaintext) and {\tt
  salary} is confidential and encrypted using AES in CBC mode with
key, $K_{salary}$. Furthermore, assume that there is a rule:
$$
\mbox{\em greater}(Employee.salary, \mbox{\em Temp}) \to \mbox{\em Boolean}
$$
Then the client would rewrite the query, thereby encrypting the
Constant 500 using a one-time key.  Furthermore, the client would pass
this one-time key and $K_{salary}$ to the TM using the program
registration protocol described below. The compiler of the UM has full
access to all meta-data (types and rules) and, thus, knows that {\tt
  department} is plaintext, that {\tt salary} is encrypted, and that
there is a rule that authorizes the TM to execute the predicate on
{\tt salary}.  Correspondingly, the UM compiles a plan that uses the
{\em Employee.department} index (if that exists) to select all IT
workers and then calls the TM to post-filter the employees that, in
addition, match the {\em salary} predicate.

There are many other ways to implement encrypted
databases using TMs.  One approach would be to implement the compiler
in the TM because query compilation can leak information, too.
Studying the affects of that approach is beyond the scope of this
paper which focuses on information leakage of the TM at the {\em running
  time} of a query. Another approach is to execute {\em all}
operations by the TM, rather than splitting the work between the UM and
the TM\@.  In this approach, the UM is merely used to do I/O, thereby
fetching data from disk for the TM, and serving as a proxy to accept
queries from clients and passing query results to the clients.  This
approach has been proposed by the TrustedDB
\cite{TrustedDB:SIGMOD2011} and Haven projects \cite{Haven:OSDI2014}.
Again, studying the confidentiality/performance tradeoffs of that
approach is beyond the scope of this paper.  Our approach are not
directly applicable to that architecture because these systems can
better constrain the capabilities of an attacker and, thus, have a
weaker attacker model (Section \ref{sec:attacks}).  However, there are
reasons why these systems are not mainstream yet and why most
start-ups and big players like Microsoft use the CryptDB-model of
encrypted databases which is also the basis for this work.  One
particular problem of systems like TrustedDB and Haven is that they
require porting an entire database engine to the TM which makes the TM
vulnerable from bugs in the DBMS engine code.
Another problem is that in the TrustedDB
and Haven architecture, it becomes difficult to administer the encrypted
database and for the cloud provider to support and debug customer
incidents.
Eventually, we believe
that both models should be studied and formalized and the main
contribution of this paper is to formalize a generalized
``CryptDB-style'' model that makes use of server-side trusted computing.  

\paragraph*{Program Registration} The rules determine which operations
(on which keys and types) the TM is allowed to perform.
Correspondingly, each authorized client that has access to encryption
keys has an authentic copy of the rule set and explicitly authorizes
the TM to execute operations using those keys. We call this
authorization process {\em program registration}.

In a nutshell, program registration works as follows.  For each rule,
the client sends a message to the TM that contains the name of the
operation and the encryption keys of all input 
and output types. This method is encrypted using the TM's public key
so that only the TM can decrypt this message and can retrieve the
keys. To this end, all TMs must be part of a public-key infrastructure
which attests that the TMs and their public keys can be trusted.

Clients carry out program registration lazily. If a function of a type
is never needed by the application, then that function is never
registered at the TM and, thus, that function is never exposed to a
potential attacker who has access to the TM\@. This lazy approach to
program registration corresponds to the onion technique proposed as
part of the CryptDB project \cite{CryptDB}. Like lazy program
registration, CryptDB degrades the level of encryption (using a weaker
PPE technique) lazily and as needed by the application by peeling off
a layer of stronger encryption from the onion. The important
difference is that this process is expensive in CryptDB because it
involves updating (re-encrypting) every tuple. In contrast, this 
process is cheap with lazy program registration because it involves
only sending a single message from a client to the TM\@. Likewise,
deregistering a function in the TM is cheap as it involves only a
single message to the TM\@. In contrast, adding a layer of encryption
to CryptDB to improve confidentiality is expensive because it again
involves updating the whole table, thereby shipping the whole table
from the server to the client, re-encrypting it at the client, and
sending the re-encrypted table back from the client to the server.

 \paragraph*{Query Optimization} The basic principles of query
 optimization for an encrypted database system are the same as for a
 traditional database system \cite{Cascades}: enumerate alternative
 plans, estimate the cost of each plan, and select the cheapest
 plan. Furthermore, the algorithms and operators used in an encrypted
 database system are the same as in traditional database systems. The
 only difference is that some  of the computation (e.g., hashing a
 value, comparisons between two  values) is carried out by the TM,
 whereas the bulk of the algorithms  and data movement are carried out
 by the UM\@.   

 Query compilation is nevertheless more complex for two
 reasons. First, the TM 
  can be seen as additional processing resources. Therefore,
 encrypted database systems with TMs require distributed
 query optimization in order to minimize the number of interactions
 between the UM and TM\@. An alternative way is to consider
 expressions that are evaluated by the TM as expensive UDFs and
 optimize queries accordingly \cite{Hellerstein:SIGMOD1993}. Second, encryption
 may limit the use of certain algorithms; for instance, hash-based
 algorithms are not applicable to Probabilistic types that represent
 the strongest level of confidentiality (Section \ref{sec:query}).
 Third, depending on the rule system, there might be no, one, or
 several ways to evaluate an expression and query using the TM\@.
 Similar issues have been studied in the context of heterogeneous
 database system with Web sources in the late Nineties
 \cite{Garlic:VLDB1997}. To
 illustrate, consider the following example query and example rules ExR1-ExR7:
 \begin{quote} 
 SELECT R.a + R.b + R.c FROM R 
 \end{quote}

 \begin{tabular}{ll} 
 ExR1: & $\mbox{\em add}(R.a, R.b) \to R.b$ \\
 ExR2: & $\mbox{\em add}(R.a, R.b) \to R.c$ \\
 ExR3: & $\mbox{\em add}(R.b, R.c) \to R.c$ \\
 ExR4: & $\mbox{\em add}(\mbox{\em Temp(Integer)}, \mbox{\em Temp(Integer)}) \to \mbox{\em Temp(Integer)}$ \\
 ExR5: & $\mbox{\em rotateKey}(R.a) \to \mbox{\em Temp(Integer)}$ \\
 ExR6: & $\mbox{\em rotateKey}(R.b) \to \mbox{\em Temp(Integer)}$ \\
 ExR7: & $\mbox{\em rotateKey}(R.c) \to \mbox{\em Temp(Integer)}$ 
 \end{tabular} 

 There are two ways to execute this query. First, with Rules ExR1 and
 ExR3. Second, with Rules ExR5, ExR6, ExR7, and ExR4. The first way
 involves two calls to the TM per record whereas the second way
 involves four calls so that the first approach is cheaper.

%% file: attacks.tex
\section{Attacker Model and Example Attacks}
\label{sec:attacks}

Why is it so difficult to build encrypted databases? Why don't trusted
computing platforms solve all problems?  The difficulty lies in
defining the right rule set and, thus, the right set of operations to
run in the TM\@.  If the wrong set of operations runs in the TM, the
system leaks information and therefore becomes insecure. 

This section defines our attacker model and presents several example
attacks that demonstrate why the rule set of an encrypted database
must be carefully chosen.  Based on these observations, the next
sections then give guidance on how to design rule sets for encrypted
databases.


\subsection{Attacker Model}
\label{subsec:attacker_model}

Looking back at Figure \ref{fig:umtm}, we assume that the client and
the TM are trusted and secure and that the attacker can fully observe
and manipulate the UM, all other components of the system, and all
communication.  Specifically, we assume that the attacker has the
following capabilities (Figure \ref{tab:attack_model}):  
\begin{enumerate}
  \item {\em Database:} The attacker can read and modify the encrypted
    database as stored on disk or in the main memory/caches of the
    UM\@. 
  \item {\em Queries:} The attacker can observe all communication
    between the UM and TM and between the UM and all clients.  In
    particular, the attacker can see queries submitted by clients and
    the (encrypted) query results returned to the clients. 
  \item {\em UM:} The attacker can read and modify the memory of the
    UM and change the code that runs in the UM\@.   
  \item {\em TM:} The attacker can call the TM as often as she wants,
    with any input.  Furthermore, we assume that the attacker knows
    which programs are registered in the TM\@. However, the attacker
    cannot read or modify the state of the TM in any way. 
\end{enumerate} 
These capabilities model a strong, active attacker.  Such a strong
attacker model helps to protect the database against many possible
attacks such as insider attacks (administrators or cloud providers),
third-party attackers (hackers who gain control of the UM,
co-tenants), and bugs in the software of the UM which is the bulk of
the database system.  In practice, the capabilities of attackers on
the UM may be more limited, but assuming such a strong attacker makes
it easier to reason about the confidentiality of the system:  The only
thing that matters are the capabilities of the TM (i.e., the functions 
supported by the TM) and the encryption technique used to protect
confidential data.

In one regard, this attacker model might be too optimistic.  It is known
that Intel SGX is subject to side-channel attacks
\cite{SGX:Explained}.  However, studying these kinds of attacks is
beyond the scope of this paper and we hope that trusted computing
technology will get better, making such attacks more difficult.  

Under this attacker model, it is easy for the attacker to launch
denial-of-service attacks. For instance, the attacker could simply
delete the database on disk or overload the TM with garbage requests.
The attacker could also drop all traffic between clients and the
database server.  Such denial-of-service attacks are outside of
the scope of this paper.  The goal of this paper is to study
confidentiality and protecting secrets stored in the encrypted
database.    

\begin{figure}
  \centerline{\small
  \begin{tabular}{|c|c|c|c|c|}  \hline
    & {\em Read State} & {\em Modify State} & {\em Observe Messages} & {\em Send Messages} \\ \hline \hline
    {\em Client} & \xmark & \xmark & \checkmark & \checkmark \\ \hline
    {\em UM} & \checkmark & \checkmark & \checkmark & \checkmark \\ \hline
    {\em TM} & \xmark & \xmark & \checkmark & \checkmark \\ \hline
  \end{tabular}
  }
  \caption{Capabilities of Attacker}
  \label{tab:attack_model}
\end{figure}

\subsection{Trivial Attacks}
\label{subsec:trivial_attack}

Let us start with a trivial example to demonstrate why the rule set of
an encrypted database must be designed carefully. The following rule
authorizes the TM to compare an (encrypted) {\em salary} with a
plaintext value:
$$
\mbox{\em equal(salary, Integer)} \to \mbox{\em Boolean}
$$
With this rule, an attacker can discover all 
salaries in the database by
calling the TM to compare them to all possible Integer values.
Likewise, the following rule which authorizes the TM to simply decrypt
an encrypted value must not be registered in the TM:
$$
\mbox{\em decrypt(salary)} \to \mbox{\em Integer}
$$

\subsection{Information Leakage from Errors}
\label{subsec:error_attack}

Another source of information leakage in an encrypted database with a
TM are errors.  Consider the following rule which authorizes the TM to
execute divisions on salaries (e.g., to implement a 10 percent salary
raise), return the result in an encrypted form, yet to expose errors
in plaintext so that the UM can react and not update a salary in the
event of an error. 
$$
\mbox{\em division(salary, salary)} \to \mbox{\em salary, Error}
$$
If the TM supports this operation, then an attacker can easily find
all employees in the database who work for free by obverving {\em
  division by zero} errors when trying to divide a salary by itself.

Exploiting overflow errors is another way for an attacker to infer
information from an encrypted database.  By observing overflow errors,
it is possible to determine which of two (encrypted) integers is
larger than the other one by iteratively muliplying these integers
with themselves until an overflow error occurs.

\subsection{Composition}
\label{subsec:enumeration_attack}

Arguably, the most dangerous attacks come from composing functionality
in the TM\@.  In this attack, each rule by itself is safe, but the
combination of rules allows the attacker to infer confidential
values. As a simple example, assume that we would like to support both
arithmetics and comparisons on salaries and define the following rules: 
\begin{quote}
{\em division(salary, salary)} $\to$ {\em salary} \\
{\em add(salary, salary)} $\to$ {\em salary} \\
{\em equal(salary, salary)} $\to$ {\em Boolean}
\end{quote}
To determine the value of a specific encrypted salary (e.g., {\em
  Bob.salary}), we execute the following steps, assuming that Bob's
salary is not 0:
\begin{enumerate}
\item EncryptedOne = TM.divide(Bob.salary, Bob.salary)
 \item {iEncrypt = EncryptedOne}; {iPlaintext = 1;} 
 \item while (NOT TM.equal(Bob.salary, iEncrypt)) 
   \begin{enumerate}
   \item iPlaintext++;
   \item iEncrypt = TM.add(iEncrypt, EncryptedOne);
   \end{enumerate}
   \item return {\em iPlaintext}
\end{enumerate}

\subsection{Other, Known Inference Attacks}
\label{subsec:inference_attacks}

All of these example attacks are specific to an encrypted
database enhanced with a trusted computing platform. In addition, all
attacks that have been studied for more traditional encrypted
databases (e.g., CryptDB) remain valid.

\cite{Seny:PPE2015} shows how to reveal confidential information from
a medical database if the data is deterministically encrypted and the
attacker exploits publicly available background knowledge about value
distributions.  This work is directly applicable to an encrypted
database with a TM, even if the data is not deterministically
encrypted.  The only thing that is needed is that the TM supports equality
({\em equal($T, T$) $\to$ Boolean} rule) and the same inference attacks
on $T$ are possible as if $T$ were deterministically encrypted.

As another example, disk access patterns can reveal
confidential information \cite{accesspatterns-ndss2012} in any kind of
encrypted database system.  For instance, if it is known that Bob gets
salary raises most frequently, then an attacker can exploit this
information to discover Bob's record in the encrypted database by
observing which record is most frequently updated. 

%% file: reasoning.tex
\section{Reasoning about Confidentiality}
\label{sec:reasoning}

The previous section gave several examples that demonstrate that it is
important to define the rule set of an encrypted database carefully.
If not, an attacker can easily infer secrets from the database.  This
section gives a framework to reason about the confidentiality of types
in an encrypted database. The key idea is to model the types of the
encyrpted database as a graph and determine potential information
leakage by analyzing this graph. 

\subsection{What is Dangerous?}
\label{subsec:dangerous}

All the example attacks of the previous section have one thing in
common: The TM returns plaintext values.  These plaintext values are
either {\em Boolean} values (Section \ref{subsec:trivial_attack}) and
inference attacks on determinic encryption \cite{Seny:PPE2015}),
errors (Section \ref{subsec:error_attack}), or pointers (inference
from access patterns \cite{accesspatterns-ndss2012}).  In other words,
when analyzing the rule set of an encrypted database, we need to look
for rules that involve plaintext types as results.  As shown in
Section \ref{subsec:trivial_attack}, the combination of plaintext
parameters and plaintext results can be particularly dangerous.

Another observation is that {\em danger} can propagate through rules.
For instance, if I add two encrypted integers and then compare the
result with another integer, then that comparison might leak
information about the first two encrypted integers.

A third observation is that rules with cycles can also be particularly
dangerous such as the {\em add} rule in Section
\ref{subsec:enumeration_attack} which allows the attacker to use the
TM to enumerate the whole domain.

\subsection{Information Flow Graph}
\label{subsec:flowgraph}

Based on these observations, we study the information leakage using an
{\em information flow graph}. The nodes of this graph are types and
edges in this graph are determined by rules. More formally, given a
schema with types $\mathcal{T}$ and rule set  $\mathcal{R}$, we define
the information flow graph $G(\mathcal{T}, \mathcal{R})$ as follows:   
\begin{enumerate}
\item Vertices are the schema types in $\mathcal{T}$.
\item There is an edge $T_1 \rightarrow T_2$ with label $f (i, j)$ if
  there is a rule involving function $f$ where $T_1$ is the $i$th
  input type of $f$ and $T_2$ is the $j$th output type of $f$. 
\end{enumerate}

The set of \emph{plaintext terminating walks} of Type $T$,
$\mathit{PTWalk} (T, \mathcal{R})$, is the (possibly infinite) set of
walks (paths) in the information flow graph ($G(\mathcal{T},
\mathcal{R})$) that start at vertex $T$ and end in a vertex
corresponding to a plaintext type.   

\begin{figure}
\centering
\includegraphics[trim=0ex 90ex 178ex 0ex, width=1.5in]{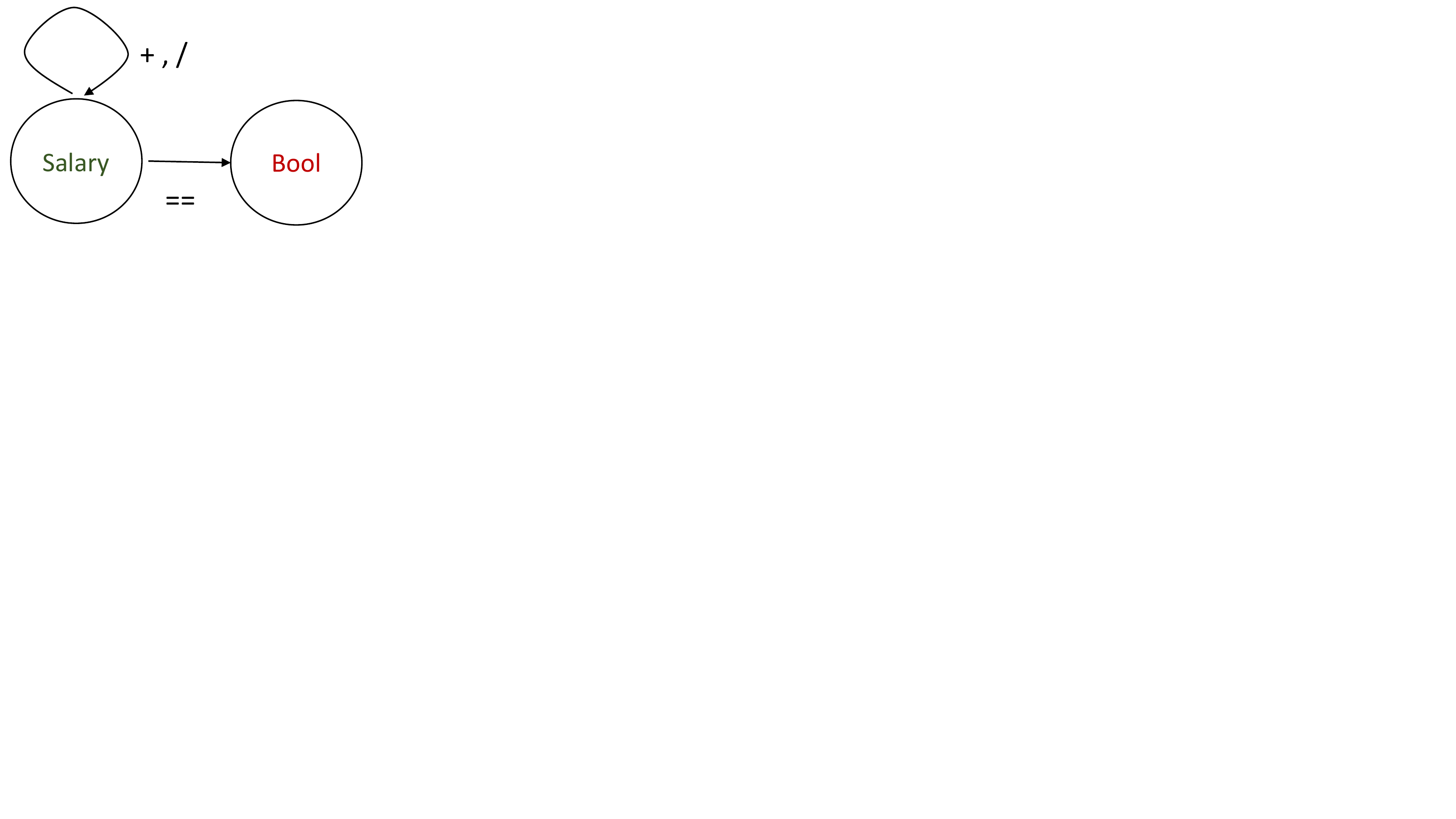}
  \caption{Example Information Flow Graph (Section
    \ref{subsec:enumeration_attack})}
  \label{fig:enumeration_attack}
  \end{figure}

With these definitions, we can now revisit the examples from the
previous section. For instance, Figure \ref{fig:enumeration_attack}
shows the information flow graph for the rule set of Section
\ref{subsec:enumeration_attack}.  Due to the cycle, there is an
infinite number of plaintext terminating walks for {\em Emp.salary}. 

\subsection{What Works?}
\label{subsec:whatworks}

A type has no information leakage if its values are encrypted using an
IND-CPA encryption technique (e.g., AES in CBC mode) and there are no
paths from this type (or any parameters of rules it is involved in) to
a plaintext type in the information flow graph.  Section
\ref{subsec:query_probabilistic} gives an example class of such types
and discusses its properties. In a nutshell, all rules follow the {\em
  encrypted in / encrypted out} principles for these types so that the
TM only takes encrypted inputs and produces encrypted outputs.

Unfortunately, as shown in Sections \ref{sec:query} and
\ref{sec:updates}, it is not always possible to follow the {\em
  encrypted in / encrypted out} principle for all types because it
limits certain SQL functionality (e.g., the use of integrity
constraints, Section \ref{sec:updates}) and the use of important
database features such as indexes (Section \ref{sec:query}).  All
these features require some paths to plaintext values.  While we
cannot prove absolute confidentiality in such situations (any path to
a plaintext type leaks information and is a potential vulnerability),
we can try to minimize the paths to plaintext values and, thus, limit
information leakage as much as possible.

\begin{figure}
\centering
\includegraphics[trim=0ex 75ex 173ex 0ex, width=1.5in]{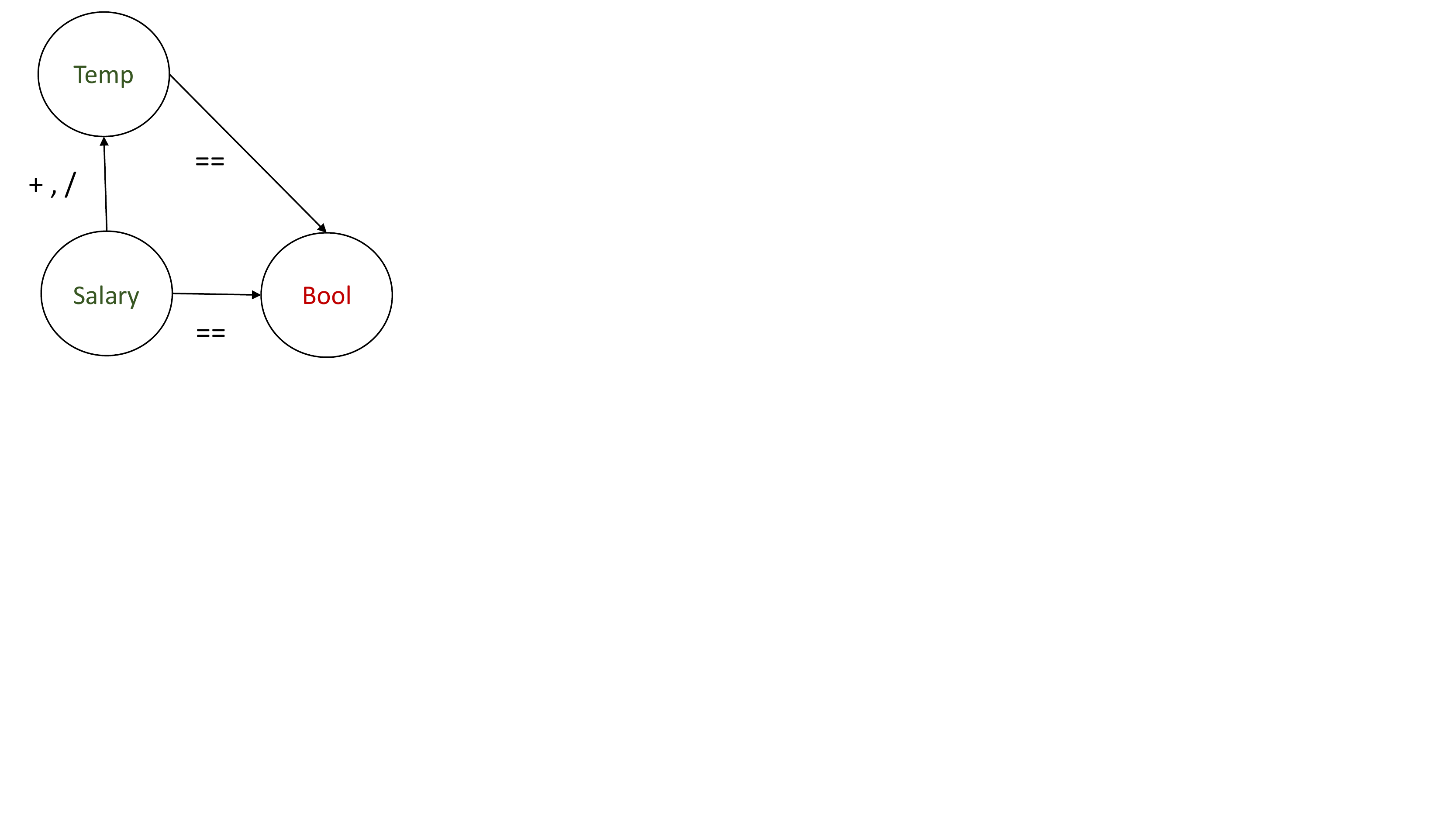}
  \caption{Breaking the Cycle}
  \label{fig:enumeration_fix}
  \end{figure}

To give an example, Figure \ref{fig:enumeration_fix} shows the
information flow graph of the following rule set that supports
arithmetics and comparisons on salaries (in particular, predicates of
the form $A+B=C$) and is not vulnerable to the enumeration attack of
Section \ref{subsec:enumeration_attack}:
\begin{quote}
{\em division(salary, salary)} $\to$ {\em Temp} \\
{\em add(salary, salary)} $\to$ {\em Temp} \\
{\em equal(salary, temp)} $\to$ {\em Boolean}
\end{quote}
The key idea of this rule set is that it breaks the cycle by using
{\em temp} types; i.e., by encrypting the result of any arithmetics
with {\em salaries} using one-time-keys. As shown in Section
\ref{subsec:query_deterministic}, this trick allows to support almost
any SQL feature without leaking more information than today's
generation of encrypted databases (e.g., CryptDB).

\subsection{Type Hierarchy}
\label{subsec:type_hierarchy}

Using the concepts of an {\em information flow graph} and {\em
  plaintext terminating walks}, it is also possible to define a
partial order on types.  A stronger type leaks as much or less
information than a weaker type.  Type $T^{R_1}$ is stronger than
$T^{R_2}$ iff $\mathit{PTWalk} (T, R1) \subseteq
\mathit{PTWalk}(T,R2)$. Here $T^{R_1}$ corresponds to schema type $T$
with rule set $R_1$ and $T^{R_2}$ to schema type $T$ with rule set
$R_2$.  Intuitively, more paths to a plaintext type mean that $R_2$ is
richer than $R_1$ and, in particular, $R_2$ supports more potentially
{\em dangerous functionality} that can result in infering values from
$T$ using plaintext values that result on computations on values of $T$.

The existence of type hierarchies makes it possible to define {\em
  levels of confidentiality}. The next sections will discuss three
possible levels.


%% file: query.tex
\section{Confidentiality vs.\ Performance Tradeoffs}
\label{sec:query}

This section defines three classes of schema types: (a)
Probabilistic, (b) Deterministic, and (c) Ordered.  We chose these
classes because they correspond to the ``confidentiality {\` a} la
carte'' offerings of state-of-the-art encrypted database systems that
are based on property-preserving encryption (PPE).  For instance,
Microsoft Always Encrypted supports probabilistic and deterministic
encryption with AES \cite{sql-AE}.  CryptDB supports
probabilistic and deterministic encryption with AES and
order-preserving encryption with a custom-defined encryption scheme
\cite{Popa:OPE2013}.

Probabilistic data types as defined here give
the same confidentiality guarantees as probabilistic encryption in
CryptDB and Always Encrypted.  Likewise, Deterministic and Ordered
types as defined here provide the same level of confidentiality
(information leakage) as deterministic and order-preserving encryption
in CryptDB\@.  The difference is that, with trusted
computing resources, we can support significantly more functionality
than Always Encrypted and CryptDB at the same level of
confidentiality.  In fact, we show that we can process any
SQL {\tt SELECT} statement at any level of confidentiality, even for
Probabilistic types.  Using a TM and our generalized model of an
encrypted database, the three levels of confidentiality differ most
significantly in performance:  The higher the level of
confidentiality, the lower the performance for certain queries.  For
instance, Probabilistic types can only make use of oblivious
algorithms and no indexes whereas indexes can be used for both
Deterministic and Ordered types. 


\subsection{Probabilistic}
\label{subsec:query_probabilistic}

\subsubsection{Definition and Confidentiality}

{\em Probabilistic} is the strongest level of confidentiality.  Given
a schema with types $\mathcal{T}$ and rules $\mathcal{R}$, a type $P
\in \mathcal{T}$ is Probabilistic if there is no path in the
information flow graph $G (\mathcal{T}, \mathcal{R})$ from $P$ to a
plaintext type.  That is, probabilistic types can only be involved in
rules in which all other types (inputs and outputs) are also
probabilistic.  Furthermore, an IND-CPA encryption algorithm must be
used for probabilistic types; e.g., AES in CBC mode.

Both Microsoft Always Encrypted and CryptDB support this level of
confidentiality, but they do not support any operations on
probabilistically encrypted data. That is, to process values of these
types, the client must download all the data (no filter possible),
decrypt the data at the client, and process any kind of query on these
values at the client.  We can show that with a server-side TM, we can
execute any SQL {\tt SELECT} statement on Probabilistic types support
and provide the same confidentiality guarantees as Always Encrypted and
CryptDB for probabilistically encrypted data.

\begin{figure}
  \centerline{\small
\framebox{
    \begin{tabular}{ll}
PR1: & $f (T_1, \ldots, T, \ldots T_n) \to T_R$ \\
PR2: & $\mathit{Pred}(T_1, \ldots, T, \ldots T_n) \to \mbox{\tt EncBool}$ \\
PR3: & $\mbox{\em identity}(T) \to T_R$ \\
PR4: & $\mbox{\em orderPair}(R_1, R_2, \mbox{\em compare}) \to R_R, R_R$ \\
\end{tabular}
  } }
  \caption{Illustration of rules consistent with Probabilistic types}
  \centerline{\small $T, T_R$ are Probabilistic and others are arbitrary types in R1-R3}
  \centerline{\small $R_1, R_2$ any Row Type; $R_R$ any Probabilistic Row Type}
\label{fig:probabilistic}
\end{figure}

\subsubsection{Functionality and Performance}

To demonstrate the expressive power of Probabilistic types, Figure
\ref{fig:probabilistic} gives example rules that are consistent with
our definition of a Probabilistic type.   The first rule, PR1,
specifies that any expression can be executed on a Probabilistic type
as long as all inputs and outputs have a Probabilistic type. The
expression, $f$, can be any SQL expression; e.g., compositions of
arithmetics, string functions, etc.  There is no need to encrypt $f$
and hide which expressions are registered in the TM\@.

As shown in Section \ref{subsec:error_attack}, all errors that might
occur as a side-effect of processing $f$ must be encrypted, too. In
PR1 (and all other example rules throughout this section), we model
errors as special values of the domain of $T_R$. When functions are
composed (e.g., the result of $f$ is used as input to another
function), then the errors are propagated; that is, the result is
again an encrypted error value.  Alternatively, encrypted errors could
also be modeled as a separate result type, as done in the example of
Section \ref{subsec:error_attack}. In this case, we would implement
composition would by providing encrypted errors as additional inputs.
These two ways of implementing compositions are equivalent.  The first
approach to model errors as special values of the encrypted result
domain is simpler and that is why we use that approach throughout the
remainder of this paper.

Depending on the {\em expression}, $f$, in Rule PR1, the condition
that {\em all} inputs must be encrypted and have a Probabilistic type
can be relaxed.  For instance, if the expression is $A + B$ and $A$ is
a secret and $B$ is not a secret (and not encrypted), then evaluating
that expression is okay if the result is encrypted and has a
Probabilistic type.  This observation is particularly important for
databases in which only few columns contain secrets. 

Another interesting way to relax Rule PR1 and extend the expressive
power is that depending on the expression, it may be okay to return
errors in plaintext.  As an example, consider the following expression
with $S$ a secret string that has a Probabilistic type: 
$$
   \mbox{\em substring}(-1, -1, S)
$$
 This expression returns an error because the position and length in
 the substring function must not be negative.  However, the attacker
 does not learn anything about $S$ from this error.  More formally,
 the error is a function of the other inputs independent of the
 (secret) string.  So, the following rule would be okay for a
 Probabilistic type, $T_S$, with domain {\em String}, if $T_R$ is also
 a Probabilistic type with domain {\em String}, and $E$ is a plaintext
 type that returns the error status.  $T_R$ may have the same or a
 different encryption key as $T_S$. 
$$
   \mbox{\em substring}(\mbox{\em Integer}, \mbox{\em Integer}, T_S) \to \langle T_R, E \rangle   
$$   
   Such a rule helps to implement updates that would otherwise not be
   possible. (Section \ref{sec:updates} discusses updates and issues
   with errors in more detail.)  However, the following rule would
   {\em not} be okay, if $T_I$ is a Probabilistic type because in this
   case plaintext errors might give away secrets of values of type
   $T_I$: 
$$
   \mbox{\em substring}(T_I, T_I, T_S) \to \langle T_R, E \rangle
$$
In general, if we wish to expose errors, we must make sure that the
output error does not depend on any input that has a Probabilistic
type.  

   The second and third rule of Figure \ref{fig:probabilistic}, PR2 and
   PR3, are special cases of PR1. They are not needed and we only list
   them here for exposition.  PR2 specifies that the TM may evaluate
   any predicate, if all inputs are Probabilistic.  Furthermore, PR2
   specifies that the result of the predicate is a probabilistically
   encrypted Boolean value so that the attacker cannot learn anything
   from the result.  An instance of {\tt EncBool} can take one of the
   following values: {\em true}, {\em false}, {\em unknown} (according
   to SQL's three-valued logic), and an (encrypted) error code, if the
   evaluation  of the predicate failed (e.g., one of the inputs of the
   predicate contained an error value).  

Like PR2, PR3 is a special case of PR1. PR3 is the {\em identity}
function that simply re-encrypts its input that is encrypted with the
key of $T_1$ using the key of $T_R$.  This identity function is
needed, for instance, to implement casts or rotate the key of a column
in the database because many organizations have a security policy that
specifies that encryption keys need to change every, say, one or two
months. PR3 is also applicable if $T_1$ is a plaintext type; PR3 is
one of the special cases of PR1 in which the inputs need not be
encrypted. This way, PR3 makes it possible to encrypt a column if a
column that was previously believed to be non-confidential becomes
confidential. 
   
  PR4 is different and critical to implement SQL queries on
  Probabilistic data in an effective way.  PR4 specifies the {\em
    orderPair()} function that takes two records as input (the fields
  of these records may be all encrypted, all plaintext, or some
  encrypted and some not encrypted), a comparison function, and
  returns the two records ordered according to the {\em comparison}
  function and fully probabilistically encrypted so that no
  information is leaked from the result.  PR4 makes sure that we can
  sort any table, thereby using BubbleSort or any other oblivious sort
  algorithm \cite{DBLP:conf/soda/Goodrich10}. Quicksort cannot be
  implemented for Probabilistic types because it requires a {\em
    partitioning} function which gives a plaintext result that
  specifies whether an element is larger or smaller than another
  element and, thus, violates the encrypted in / encrypted out
  principle. 

PR4 is defined on {\em records}, rather than values of a domain,
because the encrypted result must not give any indication of whether
the first or the second tuple is smaller; from this point on, the SQL
query processor must operate on records in which all fields are
Probabilistic to avoid any kind of inference attacks on the result of
an oblivious sort. 

In the extended version of this paper~\cite{typesystem:techreport}, we
formally prove that an adversary does not learn any instance level
information about a value of a Probabilistic type. We prove this by
setting up an indistinguishability experiment where the adversary is
not able to distinguish between two values given an oracle access to
the TM and an access to encryption oracles of other types.

\begin{theorem}(informal)
The adversary does not learn any information about a value of a Probabilistic type other than what she learns from the schema and the (encrypted) database instance (e.g., via correlations with other columns in the same table).
\end{theorem}
 
There is more good news. It turns out that the rules of Figure
\ref{fig:probabilistic} are sufficient to implement any SQL {\tt
  SELECT} query.  All database operators can be carried out using
sort-based algorithms (e.g., joins, group-by, sub-queries) and
expression evaluation.  Doing so, each tuple of the final query result
is tagged with an {\tt EncBool} value and the client can filter out
the result tuples by decrypting the tuples and this Boolean value. To
speed up, this filtering at the client, the final query result can be
sorted by the {\tt EncBool} value in the following way:  First, all
errors (if any), then all {\em true} tuples which are part of the
query result, and then all {\em unknown} and {\em false} values.  This
way, the client can immediately detect whether an error occurred and,
if not, quickly extract the result tuples.  Such oblivious query
processing has been studied extensively
in the literature; e.g., \cite{DBLP:conf/icdt/ArasuK14} for the most
recent overview.   

The bad news is that the performance of query processing on
Probabilistic types can be prohibitively poor.  The size of a join
result is quadratic with the join inputs, independent of the
selectivity of the join predicate.  Similarly, the amount of data that
needs to be shipped from the server to the client for a simple query
that applies a filter to a table (e.g., a key-value lookup) is in the
same order as the table, independent of the selectivity of the filter
predicate.  Grouping has the same problem.  Finally, oblivious RAM
simulations \cite{ORAM} are the only way to make use of indexes on
Probabilistic types; unfortunately, oblivious RAM simulations have
high overheads (e.g., poor cache locality).  In summary,
Probabilistic types are only the right choice for highly confidential
data that are rarely part of a filter predicate, join predicate, or 
group-by key, or for which performance is not a major concern.  To achieve
better performance, we need to sacrifice confidentiality.  That is
what the Deterministic and Ordered types do.   

\subsection{Deterministic}
\label{subsec:query_deterministic}

The goal of Deterministic types is to allow the encrypted database to
make the most efficient use of indexes, hash-based algorithms, and
filters to limit the size of (intermediate) query results in the same
as with a traditional, unencrypted database. Unfortunately, these
features cannot be supported without leaking some information. The
good news is that the information leakage from supporting these
features has already been studied in the context of systems like
CryptDB and SQL Always Encrypted which make use of deterministic
encryption (e.g., AES in ECB mode).  With the use of trusted
computing, we can define Deterministic types that significantly extend
the functionality that can be implemented with deterministic
encryption, thereby giving the same confidentiality guarantees. In
particular, we can expand the use of indexes to any sargable predicate
such as ``$A = B + C$'' if an index on $A$ is available; this
expansion is only possible with our specific schemes and not possible
in systems like CryptDB and Always Encrypted.  This section formally
defines Deterministic types and shows how to use indexes and
hash-based algorithms to efficiently process queries on Deterministic types.



\subsubsection{Definition and Confidentiality}

For any domain $\mathcal{D}$, let $\mathit{equal}(\mathit{D},
\mathit{D}) \rightarrow \mathit{Boolean}$ denote the equality
predicate over $\mathcal{D}$. Also, let $\mathit{hash}(\mathit{D})
\rightarrow \mathit{HashDom}$ denote a cryptographic hash function
that maps values of $\mathcal{D}$ to a plaintext hash domain; e.g.,
for SHA-1, the hash domain is the domain of 160-bit values. Let $f_1
\ldots f_k$  be injective functions; that is, $x \neq y  \implies
f_i(x) \neq f_i(y)$ for all $i = 1 \ldots k$.   

Given a schema with types $\mathcal{T}$ and rules $\mathcal{R}$, we
say a type $D$ is Deterministic if: 
\begin{enumerate}
\item All paths in $G (\mathcal{T}, \mathcal{R})$ from $D$ to $\mathit{Boolean}$ are of the form:
\[ D \overset{f_1(1,1)}{\rightarrow} T_1 \overset{f_2(1,1)}{\rightarrow} \ldots \overset{f_k(1,1)}{\rightarrow} T_k \overset{\mathit{equal}(,)}{\rightarrow} \mathit{Boolean} \]
where each $f_i$ is an injective function.
\item All paths in $G (\mathcal{T}, \mathcal{R})$ from $D$ to $\mathit{HashDom}$ are of the form:
\[ D \overset{f_1(1,1)}{\rightarrow} T_1 \overset{f_2(1,1)}{\rightarrow} \ldots \overset{f_k(1,1)}{\rightarrow} T_k \overset{\mathit{hash}(1,1)}{\rightarrow} \mathit{HashDom} \]
where each $f_i$ is an injective function.
\item For any $T_i$ in (1) and (2) above, there exists a unique path in $G (\mathcal{T}, \mathcal{R})$ from $D$ to $T_i$. 
\item There does not exist any path from $D$ to a plaintext domain that is not $\mathit{Boolean}$ or $\mathit{HashDom}$. 
\end{enumerate}

In the full version of this paper \cite{typesystem:techreport}, we
show that for any Deterministic type as defined above, an adversary
does not learn any information other than equality over its instances.  
\begin{theorem} (informal)
\label{thm:det}
Values of a Deterministic type have the same information leakage as
encrypting the values using deterministic encryption. 
\end{theorem} 

To implement a Deterministic type, we can use either probabilistic
(e.g., AES in CBC mode) or deterministic encryption (e.g., AES in ECB
mode); the choice does not impact confidentiality.  For performance
reasons, we suggest to use deterministic encryption because that way
we need less calls to the TM in order to implement comparisons and
hash values of Deterministic types. 

\subsubsection{Functionality and Performance}

To demonstrate the expressive power of Deterministic types, Figure
\ref{fig:deterministic} lists some example rules that are consistent
with the formal definition of Deterministic types. Specifically, these
rules are used to implement indexes and hash-based algorithms such as
hash joins.  Sort-based algorithms and any oblivious
algorithms that can be used for Probabilistic types can also be used
for Deterministic types.

\begin{figure}
  \centerline{\small
\framebox{
    \begin{tabular}{ll}
DR1: & $f (T_1, \ldots, T, \ldots, T_n) \to T_P$ \\
DR2: & $\mathit{Pred}(T_1, \ldots,T, \ldots T_n) \to \mbox{\tt EncBool}$ \\
DR3: & $\mbox{\em identity}(T) \to T_O$ \\
DR4: & $\mbox{\em orderPair}(R_1, R_2, \mbox{\em compare}) \to R_P, R_P$ \\
DR5: & $\mbox{\em equal}(T, T) \to Boolean$ \\
DR6: & $\mbox{\em hash}(T) \to \mathit{HashDom}$ \\
\end{tabular}
  } }
  \caption{Rules consistent with {\em Strict Deterministic Types}}
  \centerline{\small $T$ is a Deterministic Type; $T_O$
    an {\em Other} Deterministic Type} 
  \centerline{\small $T_P$ any Probabilistic Type; $R_1, R_2$ any Row Type; $R_P$ any Probabilistic Row Type}
  \centerline{\small others are arbitrary types}
\label{fig:deterministic}
\end{figure}

\paragraph*{Deterministic vs. Probabilistic Types (DR1-DR4)}

Rules DR1 to DR4 of Figure \ref{fig:deterministic} are almost the same
as Rules PR1 to PR4 for Probabilistic types (Figure
\ref{fig:probabilistic}); that is why all oblivous algorithms which
are applicable to Probabilistic types are also applicable to
Deterministic types.  Since Deterministic types support a
superset of rules as compared to Probabilistic types, it is also easy
to prove that Deterministic types are weaker than Probabilistic types
(Section \ref{subsec:type_hierarchy}).  There is, however, one subtle
point in Rules DR1, DR2, and DR4:  All the output types must be
Probabilistic.  That is, the result type of the addition of two
integers that are instances of Deterministic types is Probabilistic.
As a consequence, only the rules of Figure \ref{fig:probabilistic} are
applicable to the result of such expressions and Rules DR5-DR6 of
Figure \ref{fig:deterministic} are {\em not} applicable.  This trick
prevents the enumeration attack of Section
\ref{subsec:enumeration_attack} and uses the pattern to break cycles
discussed in Section \ref{subsec:whatworks}. 

Rule DR3 of Figure \ref{fig:deterministic} is special.  The output of
DR3 may be a Deterministic type.  Just like PR3 of Probabilistic
types, we need Rule DR3 for two purposes: (a) casts (e.g., conversion
from {\em Integer} to {\em Float}) and (b) key rotation (i.e.,
re-encrypting a column with a new key).  Casts are frequent in
practice and are not supported in any encrypted database system that
is based on PPE\@.  

\paragraph*{Strict vs. Relaxed Deterministic Types (DR5, DR6)}

DR5 and DR6 model the properties of deterministic encryption.  So,
with deterministic encryption to implement Deterministic types these
rules are not needed because these functions can just as well be
implemented in the UM in the same way as in any database system that
exploits PPE such as Always Encrypted or CryptDB\@.  One important
observation is that neither {\em equal} nor {\em hash} produce errors.
A bit trickier are null values.  These cannot be kept secret in a
Deterministic type because the comparison with a {\em null} value is
the Boolean value {\em unknown} in SQL's three-value logic. If {\em
  null} values need to be kept confidential, then Deterministic types
are not the right choice. This observation holds also for encrypted
databases that are based on PPE such as CryptDB or Always Encrypted
\cite{sql-AE}.  

Rules DR5 and DR6 of Figure \ref{fig:deterministic} model
deterministic encryption in a strict way.  With trusted computing
resources, we could also implement a more general variant of these
rules. This variant of DR5 and DR6 is shown in Figure
\ref{fig:relaxed_deterministic}.  This variant allows to compare
values of two Deterministic types with the same domain, but different
encryption keys.  In particular, it allows hash joins between two
columns that are encrypted using different keys.  (For this reason,
DR6' uses the hash function of $T_1$ to hash values of $T_2$.)  One
application for this relaxed variant is a security policy that rotates
the encryption keys of primary key columns and foreign key columns
individually.  During this rotation process the relaxed policy would
allow efficient primary key / foreign key joins (e.g., using hash
joins or index nested-loop joins) even at a moment after the primary
key column has been re-encrypted and before the foreign key column has
been re-encrypted.  Such relaxed Deterministic types could also
result, e.g., from rules $\mathit{identity}(T_1) \rightarrow T_O$ and
$\mathit{identity}(T_2) \rightarrow T_o$ where $T_1, T_2$ and $T_O$
are Deterministic types.  

On the negative side, the relaxed rule set of Figure
\ref{fig:relaxed_deterministic} might be vulnerable to additional
inference attacks by correlating (encrypted) values of $T_1$ and $T_2$
which are not supposed to be correlated.  \cite{typesystem:techreport}
contains a more detailed and formal definition of relaxed types and
discussion of the confidentiality guarantees of relaxed Deterministic
types. 

\begin{figure}
  \centerline{\small
\framebox{
    \begin{tabular}{ll}
DR5': & $\mbox{\em equal}(T_1, T_2) \to Boolean$ \\
DR6': & $\mbox{\em hash}(T_2) \to \mathit{HashDom}$ \\
DR7': & $\mbox{\em evalExpr}(T_1, \ldots, T_1, \mbox{\em expression}) \to \mbox{\em Temp}(T_1)$ \\
\end{tabular}
  } }
  \caption{Rules for {\em Relaxed Deterministic Types}}
  \centerline{\small $T_1, T_2$ any relaxed Deterministic Types, $T_1.{\cal D} = T_2.{\cal D}$}
\label{fig:relaxed_deterministic}
\end{figure}

\subsubsection{Composing Expressions and Index Lookups (DR7')}

If there is an index on column R.a, then the goal is to be able to use
this index for all sargeable predicates; i.e., predicates of the form:
$R.a = \mbox{\em expression}.$ 

If the expression is a constant or a parameter of the query, then this
constant or query parameter can be encrypted with the key of R.a and
we can use the index in a straight-forward way using any Deterministic
type (relaxed and strict).  However, for more complex expressions,
this approach no longer works.  The purpose of Rule DR7' of relaxed
Deterministic types is to evaluate such expressions and use the
index. 

As an example, let us consider the following query with a query
parameter, $:x$: 
\begin{quote}
  {\tt SELECT * FROM R, S WHERE R.a = S.a + :x;}
\end{quote}
This query cannot be executed in systems like CryptDB if R.a and S.a
are encrypted deterministically, even if R.a and S.a are encrypted
using the same key.  This query can be executed in a TM independent of
the types and keys of R.a and S.a (Probabilistic, strict
Deterministic, or relaxed Deterministic).  However, we can only use
the index on R.a, if both R.a and S.a have relaxed Deterministic types
(possibly with different encryption keys) and the rules of Figure
\ref{fig:relaxed_deterministic} are available.   DR7' is used to
evaluate the expression $S.a + :x$ by encrypting the query parameter
($:x$) using the key of S.a.  DR7', thus, corresponds to DR1 which
allows to compute any kind of arithmetics on any kind of Deterministic
type.  The important difference between DR7' and DR1 is that DR1
mandates that the result is Probabilistic whereas DR7' specifies that
the result of $S.a + :x$ is also an instance of a relaxed
Deterministic type. This way, Rules DR5' and DR6' become applicable to
the result of $S.a + :x$ and we can use the index on R.a to probe the
matching tuples. 

In order to understand the details of this approach, we need to
describe how to build and probe indexes on columns with relaxed
Deterministic types.  Rather than building the index on the encrypted
values (i.e., ciphertext) as done by Always Encrypted, we build the
index on hashes; that is, before inserting a new key into the index,
we call the TM in order to determine the hash of that key, thereby
using Rule DR6 (or DR6' which is a generalization of DR6 in this
particular case).  To probe a key, $k$, during query processing, we
apply the following approach: 
\begin{itemize}
\item Compute {\em hash}($k$). If $k$ is of type R.a, then use Rule DR6. If $k$ is of type {\em Temp}(R.a) use Rule R6'.
\item Use the index to lookup {\em hash}($k$).
  \item For all matches $i$ with {\em hash}($i$) = {\em hash}($k$), check whether $k = i$. If the type of $k$ is R.a, then this test can be done directly without using the TM\@.  If the type of $k$ is {\em Temp}(R.a), then call the TM to do this comparison (Rule DR5'). 
\end{itemize}

Why does this approach prevent the enumeration attack?  The key idea
is to use a new, different encryption key to encrypt the result of an
expression.  This way, the cycle is broken which is needed to
enumerate all values of a domain in the enumeration attack.   



\subsection{Ordered}
\label{subsec:query_ordered}

The last and lowest level of confidentiality that we would like to
discuss is {\em Ordered}.  We model Ordered types using
order-preserving encryption (OPE) as a role model.  Order-preserving
encryption has been studied extensively in the past \cite{Agrawal:SIGMOD2004,
  Boldyreva:OPE2012, Sanamrad:DBSec2014, Popa:OPE2013}.  It is controversial whether it is
strong enough for practical applications.  We only cover it here for completeness and without
taking a stand on its practicality.

The goal of OPE is to provide efficient implementations of range
predicates and Top N queries; in particular, using ordered indexes
such as B-trees. Ordered types can be formalized by using the
$\mathit{compare} (\mathcal{D}, \mathcal{D}) \rightarrow \{-1, 0, 1
\}$ instead of $\mathit{equal}$ in the definition of Deterministic; we
defer the details to the full version of this paper
\cite{typesystem:techreport}.  

Within our framework, Ordered types can be implemented by enabling the
following additional rule in addition to all the rules for
Deterministic types (Figure \ref{fig:deterministic}): 
$$
\mbox{\em compare}(T_1, T_2) \to \{-1, 0, 1\}
$$
By supporting this additional rule, it is obvious that Ordered types
are weaker than Deterministic types which are in turn weaker than
Probabilistic types (Section \ref{subsec:type_hierarchy}). 

As an encryption technique, any OPE technique or deterministic
encryption technique or even probabilistic technique can be used.  In
terms of confidentiality and functionality, the choice of the
encryption technique does not matter, assuming that the crypto is
strong and cannot be broken.  However, OPE achieves best performance
because it avoids calls to the TM, in the same way as deterministic
encryption achieves the best performance for Deterministic types
(Section \ref{subsec:query_deterministic}).

\subsection{Discussion}
\label{subsec:query_summary}

This section described three levels of confidentiality.  All three
levels allow the implementation of any SQL {\tt SELECT} statement on
encrypted data.  These three
levels, however, differ in terms of confidentiality and performance.
In terms of confidentiality, these levels correspond to the levels of
confidentiality supported in state-of-the-art encrypted database
systems using PPE; i.e., the lower levels of confidentiality,
Deterministic and Ordered, are vulnerable to certain kinds of
inference attacks \cite{Seny:PPE2015} whereas Probabilistic types are
semantically secure.  In terms of performance, a lower level of
confidentiality can improve performance in two ways: (a) It allows the
use of a wider range of algorithms and index structures; (b)
(intermediate) result sizes are smaller. 

Figure \ref{tab:levels_algos} shows which kind of algorithms can be
applied to which kind of data. Using a TM, scan-based and sort-based
algorithms (e.g., nested-loop joins, sort-merge joins) can be applied
to any kind of data. Hash-based algorithms (e.g., hybrid-hash joins),
however, can only be applied to Deterministic and Ordered types.
Likewise, only Deterministic and Ordered types can make use of
indexes.  Figure \ref{tab:levels_sizes} shows the size of encrypted
results, depending on the operation and the level of confidentiality.
For instance, the size of the encrypted result of a filter is of the
same order as the size of the base table (denoted as $\mathit{O(n)}$
in Figure \ref{tab:levels_sizes}) as explained in Section
\ref{subsec:query_probabilistic}.  Obviously, this results in poor
performance if further operations such as joins are applied to the
result of the filter or large result sets must be shipped to the
client.  However, queries that apply an aggregate (e.g., {\em count})
after the filter can be processed fairly efficiently even on
Probabilistic types.

\begin{figure}
\centerline{\small
\begin{tabular}{|c|c|c|c|} \hline
 & {\em Probabilistic} & {\em Deterministic} & {\em Ordered} \\ \hline \hline
Nested-loop Algorithms & \checkmark & \checkmark & \checkmark \\ \hline
Sort-based Algorithms & \checkmark & \checkmark & \checkmark \\ \hline
Hash-based Algorithms & \xmark & \checkmark & \checkmark \\ \hline
Equality Indexes & \xmark & \checkmark & \checkmark \\ \hline
Ordered Indexes & \xmark & \xmark & \checkmark \\ \hline
\end{tabular}
}
\caption{Confidentiality Levels and Permissible Algorithms}
\label{tab:levels_algos}
\end{figure}

\begin{figure}
\centerline{\small
\begin{tabular}{|c|c|c|c|} \hline
 & {\em Probabilistic} & {\em Deterministic} & {\em Ordered} \\ \hline \hline
 Aggregation & $\mathit{O(1)}$ & $\mathit{O(1)}$ & $\mathit{O(1)}$ \\ \hline
 Equi Joins & $\mathit{O(max(n_1, n_2))}$ & $\mathit{O(r)}$ & $\mathit{O(r)}$ \\ \hline
 Theta Joins & $\mathit{O(n_1 * n_2)}$ & $\mathit{O(n_1 * n_2)}$ & $\mathit{O(n_1 * n_2)}$ \\ \hline
 Filter (Equality) & $\mathit{O(n)}$ & $\mathit{O(r)}$ & $\mathit{O(r)}$ \\ \hline
 Filter (Range) & $\mathit{O(n)}$ & $\mathit{O(n)}$ & $\mathit{O(r)}$ \\ \hline
\end{tabular}
}
\caption{Confidentiality Levels and Result Sizes}
\centerline{\small $n, n_1, n_2$ size of input tables; $r$ size of result table}
\label{tab:levels_sizes}
\end{figure}
 
Overall, these results confirm the importance of the confidentiality
{\` a} la carte approach that allows users to chose the right type for
each column depending on the workload and confidentiality
requirements.  This is particularly important for enterprise workloads
and benchmarks such as TPC-C; for instance, it is typically affordable
from a performance perspective to use a Probabilistic type for credit
card numbers whereas it is good to use a Deterministic type for order
numbers because inference attacks are difficult for order numbers as
the attacker likely does not have much background knowledge on order
numbers.  Confidentiality {\` a} la carte can also be offered with
systems that are based on PPE only such as CryptDB and Always
Encrypted.  However, these systems severely limit the kinds of
functions that can be applied to encrypted data, whereas all SQL {\tt
  SELECT} statements are executable with trusted computing resources,
regardless of the level of confidentiality.

%% file: updates.tex
\section{Confidentiality vs.\ Functionality}
\label{sec:updates}

The previous section showed that any SQL {\tt SELECT} statement can be
executed on encrypted data and that the performance depends on the level of
confidentiality.  In some sense, all that was good news providing a
nice security / performance tradeoff for SQL {\tt SELECT} statements
on encrypted data. This section shows contains some bad news
concerning SQL {\tt UPDATE} statements. It contains three example
corner cases that demonstrate that some SQL updates can only be
executed with lower levels of confidentiality (limiting the
functionality of Probabilistic types), that for some updates with
predicates a lower level of confidentiality does not help to improve
performance (limiting the effectiveness of weaker levels of
confidentiality), and that some updates cannot be implemented with any
of the three confidentiality levels presented in the previous section
(limiting the functionality of any kind of encrypted database).  


\subsection{Integrity Constraints}

Integrity constraints can only be implemented on Deterministic and
Ordered types. This is because checking an integrity constraint
implicitly carries out a comparison with a plaintext result.

As an example, consider a database with order and customer
information.  {\em name} is the key of the {\em Customer} table and
every order refers to a customer (with a foreign key on {\em name}).
Furthermore, let us assume that customer names are confidential so
that {\em Order.customer} and {\em Customer.name} have a Probabilistic
type.  If there were a {\em unique} or {\em primary key} constraint on
{\em Customer.name}, then an attacker could infer whether an order
refers to a specific customer or two orders refer to the same customer
by deleting and inserting tuples into the {\em Customer} table.  To
find out whether two orders, $O_1$ and $O_2$, refer to the same
customer (and thus possibly launch an inference attack on the {\em
  Order} table, the attack could issue the following sequence of SQL
statements.
\begin{itemize}
\item DELETE FROM Customer;
\item INSERT INTO Customer(name) VALUES ($O_1$.name);
\item INSERT INTO Customer(name) VALUES ($O_2$.name);
\item ABORT;  // rollback all updates, restore Customer table
\end{itemize}
If there is a {\em unique} constraint on {\em Customer.name} and the
second {\tt INSERT} fails, then the attacker can infer that {\em
  $O_1$.name == $O_2$.name} which makes {\em Order.name} a
Deterministic type, rather than a Probabilistic type as required by
the database designer. 

Due to this kind of information leakage, it is critical that an
attacker does not have the authority to change any meta-data of the
database. Only the owner of the database is allowed to create
integrity constraints and change the schema of an encrypted database. 

\subsection{Updates}
\label{subsec:update_problem}

Consider the following SQL update statement that gives all employees
of a certain salary class a 10 percent raise:
\begin{quote}
  {\tt UPDATE Emp SET salary = 1.1 * salary WHERE salary = 100;}
\end{quote}
If {\em Emp.salary} has a Probabilistic type, the right way to execute
this update statement is to register the following program at the TM
(using Rule PR1 of Figure \ref{fig:probabilistic}):
\begin{quote}
  {\em raiseIf100(salary)} $\to$ {\em salary} \\
  \\
  {\tt newsalary = decrypt($K_{salary}$, input);} \\
  {\tt if (newsalary) == 100) newsalary = 1.1 * newsalary;} \\
  {\tt return encrypt($K_{salary}$, newsalary);}
\end{quote}
This program is called for every employee and the salary of all
employees will be updated, even if unchanged. Note that an IND-CPA
encryption technique for a Probabilistic type generates a different
ciphertext for the same plaintext so that an attacker does not know
which salaries were updated.

With a Deterministic type, one would expect that only the salaries of
employees with salary == 100 would be updated.  That is, the
expectated plan would be to select all employees that match the {\tt
  WHERE} clause (using Rule DR5 of Figure \ref{fig:deterministic}) and
then to update those employees using Rule DR1 of Figure
\ref{fig:deterministic} to compute the new salary.  Unfortunately,
however, this plan is not legal because Rule DR1 cannot be used to
compute the new salary.  The reason is that the results of Rule DR1
must have a Probabilistic type in order to break the cycle as
described in Section \ref{subsec:whatworks}.  In this example,
however, we assumed that {\em Emp.salary} has a deterministic type.

So, unfortunately, a Deterministic {\em Emp.salary} type does not help
to improve the performance of this update statement. Even if {\em
  Emp.salary} has a Deterministic type, this update needs to be
processed in the same way as if {\em Emp.salary} had a Probabilistic
type.  In fact, it is even worse because we need to generate a new key
to re-encrypt the salary column (using Rule DR3) because of the
inapplicability of Rule DR1.

This example nicely demonstrates how subtle and tricky it is to implement
encrypted databases and why secure hardware does not solve all
problems.

\subsection{Updates and Errors}

Let us revisit the update statement from the previous subsection that
raised the salaries of employees.  In fact, we made an important
assumption in the previous subsection: the expression ``1.1 * salary''
does not return any errors. It turns out that if this assumption does
not hold, then this update statement is not implementable using any
level of confidentiality.  The problem is that errors must be
encrypted in all levels of confidentiality to avoid the simple attacks
described in Section \ref{subsec:error_attack}.  While encrypting and,
thus, hiding errors is possible while processing {\tt SELECT}
statements (the error can be caught in the driver at the client),
errors cannot be hidden while processing update statements (including
deletes and inserts).  If an update occurs while processing an update,
the whole update statement must fail and this failure cannot be hidden
from an attacker.

This example shows that it is not possible to implement the whole SQL
standard on top of an encrypted database even for low levels of
confidentiality and even if we have perfectly secure trusted computing
technology.

%% file: related.tex
\section{Related Work}
\label{sec:related}

There has been a great deal of prior work on encrypted
databases. Almost all existing systems take a particular point in the
confidentiality / functionality / performance design space. One
prominent example is CryptDB\@, which is based on property preserving
encryption \cite{CryptDB}. Because of this, CryptDB is limited to the
specific confidentiality / functionality / performance characteristics
of state-of-the-art PPE techniques.  As a result, the supported
functionality is severely limited in terms of Probabilistic and
Deterministic types.  In the system suggested here, the use of
server-side TMs opens up new opportunities to implement SQL
functionality on encrypted data that cannot be implemented in CryptDB
and to achieve better performance at the same of better levels of
confidentiality. 

One particular innovation of CryptDB is the use of the {\em onion
   technique}.  In CryptDB, peeling off a layer of confidentiality
requires updating an entire column, potentially making the whole table 
unavailable during this process.  Adding a layer also requires
shipping all the data to and from the client for re-encryption.  The
proposed program registration scheme of this paper (Section
\ref{subsec:implementation}) has the same advantages of CryptDB's onion
technique (dynamically increasing and reducing the confidentiality
level) without paying the high price: Using a server-side TM, we can
add or remove a layer of protection from a column by simply
deactivating and activating TM functionality (i.e., rules).    

The closest related work is previous work on encrypted databases that
makes use of secure hardware; e.g., TrustedDB
\cite{TrustedDB:SIGMOD2011}, Haven \cite{Haven:OSDI2014},
Cipherbase \cite{Cipherbase:CIDR2013, Cipherbase:ICDE2015}, and VC3
\cite{VC3:IEEE2015}.  The result of this work are not directly applicable to
TrustedDB and Haven because TrustedDB and Haven have a different
attacker model (Section \ref{subsec:attacker_model}).  In TrustedDB
and Haven, attackers can be prevented from executing queries on the TM
by special authorization techniques.  This property, however, comes at
a cost:  The whole database must be encrypted and the whole database
engine must be ported and run inside the TM which results in
vulnerabilities from bugs and potential performance issues.  It is an
interesting and important avenue of future work to formalize and study
the confidentiality and corresponding tradeoffs of systems like
TrustedDB and Haven once these systems have matured and become
available.  To date, these systems are still research prototypes and
it is difficult to formalize their confidentality properties as these
properties are still unclear.

Our results are directly applicable to the Cipherbase and VC3 projects
which have the same attacker model as CryptDB and all other systems
that make use of PPE\@.  The main contributions of the Cipherbase 
are novel optimization techniques to execute queries and update
transactions on encrypted databases with a server-side TM\@.  These
techniques are directly applicable to our work, too, and orthogonal to
the results of this paper. 

This work is based on a large body of work that has explored inference
attacks from PPE (e.g., \cite{Seny:PPE2015}), inference attacks from
access patterns (e.g., \cite{accesspatterns-ndss2012}), oblivioius RAM
(e.g., \cite{ORAM}), and oblivious algorithms (e.g.,
\cite{DBLP:conf/icdt/ArasuK14}).  Our main 
contribution is to extend that work and come to a more general way to
reason about information leaks and cleanly characterize the
confidentiality/functionality/performance tradeoffs of
state-of-the-art encrypted database systems with secure hardware.

The holy grail of encrypted databases is fully homomorphic encryption
(FHE) \cite{Gentry:FHE2009}. With FHE, this work would indeed become
obsolete and trusted hardware would not be required. Unfortunately, we
are still a far cry away from practical FHE\@. 

%% file: concl.tex
\section{Conclusion}

This paper lays the foundation for a ``confidentiality {\' a} la
carte'' encrypted database system that make use of server-side trusted
computing. It studied the fundamental limits of those systems and
defined three levels of confidentiality that were inspired by today's
generation of property-preserving encryption techniques (PPE).  We
showed that compared to PPE-based systems we can achieve the same
confidentiality, but with much higher functionality and potentially
better  performance because we support more algorithmic variants
(e.g., sort-based algorithms for Probabilistic types).  Nevertheless,
this work confirmed previous results that there cannot be a {\em
  perfect} encrypted database system that provides semantically
security, an implementation of the full SQL standard, and good
performance.  That is why ``confidentiality {\' a} la carte'' will
continue to be the right model for encrypted databases.
 
There are several avenues for future work. First, the same basic
principles laid in this work apply to problems such as directory
services (e.g., LDAP, Microsoft Active Directory), collaboration tools
(e.g., Wikis, Sharepoint), and any other structured workload where
expression evaluation can be factored out.  Database systems are just
one example of systems that store and process confidential data.
Second, there is still a great deal of research required to implement
more efficient trusted computing-based database systems:
(Distributed) query optimization needs to be revisited and the
architecture of Figure \ref{fig:umtm} inspires research on new
algorithms to carry out joins, aggregation, and sorting efficiently
between an untrusted and trusted machine. At the moment, this work is
still at the conceptual level and we will publish the results of
performance experiments in a future paper.